\documentclass[11pt]{article}

\begin{document}

\begin{center}
{\Large \bf On the Systematic Errors in the Detection of the
Lense-Thirring Effect with a Mars Orbiter}\\
\end{center}

\begin{centering}
Giampiero Sindoni, Claudio Paris and Paolo Ialongo\\
\end{centering}
\vspace{.25in}

\begin{centering}
``Sapienza'', Universit\`{a} di
Roma, Roma, Italy\\
\end{centering}

\vspace{80pt}

We show here that the recent claim of a test of the Lense-Thirring
effect with an error of 0.5 $\%$ using the Mars Global Surveyor is
misleading and the quoted error is incorrect by a factor of at least
ten thousand. Indeed, the simple error analysis of \cite{iorio}
neglects the role of some important systematic errors affecting
the out-of-plane acceleration. The preliminary error analysis
presented here shows that even an optimistic uncertainty for this
measurement is at the level of, at least, $\sim 3026 \%$ to $\sim
4811 \%$, i.e., even an optimistic uncertainty is about 30 to 48
times the Lense-Thirring effect. In other words by including only
some systematic errors we obtained an uncertainty almost ten thousand
times larger than the claimed 0.5 $\%$ error.

\section{Systematic Errors in the Measurement of the Lense-Thirring
Effect with a Mars Orbiter: Introduction}

Before discussing the error analysis related to the strange claim of
the measurement of the Lense-Thirring effect with the Mars Global
Surveyor discussed in \cite{iorio,iorioNS}, we stress that the
statement in section (1) and in the abstract of \cite{iorio}, that
the measurement of the Lense-Thirring effect with the LAGEOS
satellites ``is still controversial'' is simply wrong and
misleading, in fact all the literature quoted by Iorio, apart from
the literature signed by Iorio only, fully confirms the $10 \%$
error budget of this LAGEOS measurement. 

On other hand, even the papers by Iorio simply claim that the error budget should have been
twice larger. Therefore even if one considers the claims of Iorio to
be right, claims that however have been shown to be wrong in several
detailed papers not only by the authors of \cite{ciupav}, the error
would be of about 20 $\%$ and of course this still is a {\it
measurement} of the Lense-Thirring effect.

However, we show here that having some familiarity with
error analysis and with data analysis, one can derive an error
budget that differs from Iorio's error budget by at least a factor of
about ten thousand!

Furthermore, anybody can repeat the LAGEOS data analysis and the
details of this LAGEOS analysis are clear to everybody since they
have been published in a number of detailed papers, see, e.g.,
\cite{ciupavper}; indeed, in a forthcoming paper it is shown how the
results of different, independent, groups do confirm  the previous
2004 measurement with the LAGEOS satellites. On the other hand, the
analysis of other experiments such as {\bf Gravity Probe B} could
hardly be understood or repeated. 

Finally, Iorio forgets to mention the relevant literature that led
to the measurement of the Lense-Thirring effect using the LAGEOS
satellites (see, e.g., I. Ciufolini, Phys. Rev. Lett., 1986
\cite{ciu1}; Int. Journ. of Mod. Phys., 1989 \cite{ciu2}; and Il
Nuovo Cimento A, 1996 \cite{ciu3}).

In section (1) of \cite{iorio} the equation $\Delta N = a (1 +
{{e^2} \over 2})^{0.5} \; sin \, i \; \Delta \Omega$ is displayed,
this equation can be found in any text book of celestial mechanics and
relates the shift of the nodal longitude $\Delta \Omega$ with the
out-of-plane shift $\Delta N$. Therefore, any serious discussion of
any measurement using this equation cannot avoid a treatment of all
the errors that enter in the modeling of $\Delta N$, i.e., a
treatment of all the nodal uncertainties {\it must} be provided.

In the following, we just treat some of the main sources of error in
the modeling of the out-of-plane acceleration. The orbital
perturbations that we treat here are: (a) perturbations due to the
main spherical harmonics of the Mars gravity field; (b) errors in
the knowledge of the orbital parameters of the Mars Global Surveyor;
(c) uncertainties in the modeling of solar radiation pressure and
planetary radiation pressure. A discussion about the other error
sources may be the subject of a following, more detailed, paper.

Here we shall compare directly the uncertainties in the nodal rates.
Translating this figures into the mean normal shift $\Delta N$ is
just a trivial step.

\section{Preliminary Error Analysis}

\subsection{Uncertainties in the gravitational perturbations}

The nodal drift due to the even zonal harmonics is:

\begin{displaymath}
\dot \Omega ^{Class} \, = - {3 \over 2} \, n \, \left ( \, {R_\oplus
\over a} \right )^2 \, {cos \, I \over \left ( \, 1 - e^2 \, \right
)^2 } \, \Biggl \{ \, J_2 + J_4 \, \Biggl [ \, {5 \over 8} \, \left
( \, {R_\oplus \over a} \, \right )^2 \, \times
\end{displaymath}

\begin{equation}
\times ( \, 7 \, sin^2 \, I - 4 \, ) \, {\left ( \, 1 + {3 \over 2}
\, e^2 \right ) \over \left ( \, 1 - e^2 \, \right )^2 } \, \Biggr ]
+  \Sigma \, N_{2n} \times J_{2n} \, \Biggr \}
\end{equation}

In the case of the Mars gravity field, a state of the art model is
for example the GGM1041c model \cite{field1}. In this recent,
accurate, model we find that the value of the quadrupole coefficient
$C_{20}$ of Mars is $-8.7450461309664714 \times 10^{-4}$, where its
formal uncertainty is $8.6998585172904000 \times 10^{-11}$. Even
though this uncertainty may not include systematic errors we shall
use it for our preliminary estimate of the total error. The
corresponding uncertainty would however be optimistic; for example
if we take the difference between the second degree even zonal
coefficient of this Mars gravity field model and the one of the
other recent Mars gravity model GMM-2B \cite{field2}, we find
$8.57721772895559 \times 10^{-10}$, i.e., an uncertainty ten times
larger; this last figure might indeed reflect the systematic errors
in $C_{20}$.

Similarly the GGM1041C value of the even zonal coefficient of Mars
of degree four, $C_{40}$, is: $5.1227082083113746 \times 10^{-6}$,
however its $formal$ uncertainty is: $7.6882432484647995 \times
10^{-11}$. Once again this is just a formal uncertainty but we shall
 use it in our optimistic error analysis.

By then inserting these two uncertainties in the equation (1) we
find a nodal rate error of about 63 milliarcsec/yr due to the
uncertainty $\delta C_{20}$ and a nodal rate error of about 112
milliarcsec/yr due to $\delta C_{40}$.

We have further calculated the uncertainty only up to degree ten,
i.e., we have only included the first 5 even zonal harmonics
uncertainties, once again this will produce a very optimistic error
budget since the MGS spacecraft is orbiting at only about 400 km of
altitude from the Mars surface and therefore higher even zonal
harmonics uncertainties would be very important in the final error
budget.

By {\bf only} including in our calculation the first 5 even zonal
harmonics, we found, in the hypothesis of non-independent errors, a total uncertainty of 707 milliarcsec/yr. Once again
this figure is very optimistic since it neglects the large errors
due to the harmonics of degree $> 10$.

Now, by taking the Lense-Thirring nodal rate of 33 milliarcsec/yr calculated by Iorio, by neglecting the important errors due to higher even zonal harmonics uncertainties and by just using
the formal uncertainties to estimate the nodal rate error (due to
the uncertainty in the even zonal harmonics up to degree 10) we then
have an error 21.4 times the size of the Lense-Thirring effect, that is:\\

{\bf Error due to First 5 Mars Even Zonals = 2140 $\%$ the Lense-Thirring effect} \\

{\bf Uncertainties in the GM of Mars}

In \cite{kono} several values of the GM of Mars, obtained with
different techniques, are reported. From these values, it is
possible to infer a relative uncertainty in the GM of Mars of about
$2 \times 10^{-7}$. By plugging this uncertainty in
equation (1) we then find an uncertainty in the nodal drift of MGS
of about 66 milliarcsec/yr, that is about:\\

{\bf Error due uncertainty in GM of Mars = 200 $\%$ the Lense-Thirring effect} \\

\subsection{Uncertainties in the Mars Global Surveyor orbital parameters}

In \cite{iorio} an uncertainty of 15 cm in the semimajor axis of the
Mars Global Surveyor is assumed. Even by assuming this uncertainty
to be realistic, by plugging it in equation (1), describing the
nodal drift of the Mars orbiter, we find an uncertainty of about 88
milliarcsec/yr, i.e.,\\

{\bf Error due to uncertainty in semimajor axis of MGS = 267 $\%$
the Lense-Thirring effect}\\

Next, we estimate the error due to the uncertainty in the
inclination of the Mars Global Surveyor. In order to provide a
figure for this uncertainty, we assume an error in the inclination
that at the MGS altitude is comparable in size with the error in the
semimajor axis, i.e., 15 cm. By thus considering an error of 15 cm
this translates into an error of about $3.96 10^{-8}$ rad in the
inclination. Finally, by plugging this uncertainty in equation (1)
we find an error of about 504 milliarcsec/yr; this large uncertainty
is due to the quasi-polar orbit of the Mars Global Surveyor which
makes it very sensitive to any small change or modeling error in the
inclination. We then have:\\

{\bf Error due to uncertainty in inclination of MGS = 1527 $\%$ the
Lense-Thirring effect}\\

For simplicity, we have neglected here the error due to the
uncertainty in the eccentricity of MGS.\\

{\bf Preliminary calculation of the uncertainty in modeling solar
and planetary radiation pressure}

In \cite{lemo2}, see also figure (3) of \cite{lemo2}, the size of
the solar radiation pressure acceleration on the Mars Global
Surveyor is reported to be equal to about $7.1 \times 10^{-6}
cm/s^2$; the planetary radiation pressure to be about $8.4 \times
10^{-7} cm/s^2$ and the MGS atmospheric drag $3.8 \times 10^{-7}
cm/s^2$. On other hand, the Lense-Thirring acceleration is on MGS of
the order of $10^{-9} cm/s^2$. Since the MGS spacecraft, contrary to
the simple spherical shape of the LAGEOS satellites, has a complex
shape, the modeling of the non-gravitational accelerations is a
rather difficult task. Even by assuming a 10 $\%$ uncertainty in the
modeling of the out-of-plane radiation pressure effects on MGS, even
by optimistically neglecting the atmospheric drag on the MGS node
due to its density variations and finally even by assuming an
optimistic reduction of the overall nodal drift due to solar
radiation pressure by a factor 0.0085, i.e., by a factor equal to
the MGS eccentricity, and thus by neglecting phenomena such as
spacecraft eclipses and thermal drag, we have that the unmodeled
nodal acceleration due to solar and planetary acceleration is $0.1
\times 0.0085 \times (7.1+0.84) \times 10^{-6} cm/s^2$, that is
$6.749 \times 10^{-9} cm/s^2$. Even though a precise calculation can
be performed following the analysis of \cite{ciu3,lucc} it is clear
that the size of the radiation pressure
unmodeled effects is then at least:\\

{\bf Error in solar and planetary radiation
pressure = 675 $\%$ the Lense-Thirring effect}\\

\section{Conclusions}

In conclusion, for the gravitational perturbations we have:\\

\noindent {\bf Error due to Gravitational Perturbations = 2342 $\%$ the Lense-Thirring effect} \\

\noindent for the uncertainties in the orbital parameters, we have:\\

\noindent {\bf Error due to MGS orbital parameters uncertainties  = 1794 $\%$ the Lense-Thirring effect} \\

\noindent and for the uncertainties in the radiation pressure
effects, we  have:\\

\noindent {\bf Error in solar and planetary radiation pressure = 675 $\%$ the Lense-Thirring effect}\\

{\bf If we sum these errors we get an error of about 4811 $\%$.
Even by taking the root-sum-square of these uncertainties we finally
have a RSS error on MGS of 3026 $\%$ of the Lense-Thirring effect!
However, it must be noted that this figure is very optimistic since
we have neglected a number of large perturbations such as the
systematic errors in the uncertainties of the first five even zonal
harmonics and the errors in the harmonics of degree higher than 10!}

Considering that in \cite{iorio} is reported an error of about 0.5
$\%$ of the Lense-Thirring effect, our optimistic estimate of the
real error is a factor about ten thousand larger than what quoted in
\cite{iorio}.


\begin{thebibliography}{99}

\bibitem{iorio} L. Iorio, High-precisionmeasurement of frame-dragging with the Mars Global
Surveyor spacecraft in the gravitational field of Mars,
ArXiv:gr-qc/0701042v5, 2007. See also L. Iorio, Class. Quantum
Grav., {\bf 23}, 5451-5454 (2006).

\bibitem{iorioNS} ``Loner stakes claim to gravity prize'', New Scientist magazine, 18 January 2007, p. 2587.

\bibitem{ciupav} I. Ciufolini, E.C. and Pavlis, , Letters to Nature, {\bf 431}, 958, 2004.

\bibitem{ciupavper} I. Ciufolini I, E.C. Pavlis and R. Peron, New
Astronomy, {\bf 11}, 527-550 (2006).


\bibitem{ciu1} I. Ciufolini, {\it Phys. Rev. Lett.} {\bf 56} 278,
1986.

\bibitem{ciu2} I. Ciufolini, {\it Int. J. Mod. Phys. A} {\bf 4}
3083, 1989. See also: Tapley B, Ciufolini I, Ries J C, Eanes R J,
Watkins M M, {\it NASA-ASI Study on LAGEOS III}, CSR-UT publication
n. CSR-89-3, Austin, Texas, 1989; See also I. Ciufolini, A. Paolozzi
et al. {\it LARES phase A study for ASI}, 1998 and {\it WEBER
SAT/LARES Study for INFN}, 2004.

\bibitem{ciu3} I. Ciufolini, {\it Nuovo Cimento A} {\bf 109} 1709,
1996.


\bibitem{field1} Lemoine, F.G., MGM1041c Gravity Model, Mars Global Surveyor Radio
Science Archival Volume MGS-M-RSS-5-SDP-V1/mors 1021, URL:
http://pds-geosciences.wustl.edu/geodata/mgs-m-rss-5-sdp-v1/mors
1021/sha/, Geosciences Node, Planetary Data System, Washington
University, St. Louis, Missouri, March 28, 2003.


\bibitem{field2} Lemoine F.G., Smith D.E., Rowlands D.D., Zuber M.T., Neumann G.A.,
Chinn D.S., and Pavlis D.E., An Improved Solution of the Gravity
Field of Mars (GMM-2B) From Mars Global Surveyor," J. Geophys. Res.,
Planets, {\bf 106} (E10), 23359-23376, 25 Oct. 2001.

\bibitem{lemo2} F. Lemoine, S. Bruinsmay, D. S. Chinnz, J. M. Forbesx, AIAA/AAS Astrodynamics
Specialist Conference, August 21-24, 2006, Keystone, Colorado, 2006.

\bibitem{kono} Alex S. Konopliv ., Charles F. Yoder, E. Myles Standish, Dah-Ning Yuan, William L.
Sjogren, Icarus {\bf 182} 23–50, 2006.

\bibitem{lucc} D.M. Lucchesi, P\&SS {\bf 50} 1067, 2002.



\end{thebibliography}
\end{document}